\documentclass[10pt]{JHEP3}
\pdfoutput=1
\usepackage{amsmath}
\usepackage{verbatim}
\usepackage{graphicx}

\newcommand{\e}{\epsilon}

\renewcommand{\L}{{\mathcal{L}}}
\newcommand{\bL}{\bar{{\mathcal{L}}}}

\newcommand{\be}[1]{ \begin{equation}\label{#1} }
\newcommand{\ee}{\end{equation}}
\newcommand{\ben}[1]{\begin{eqnarray}\label{#1} }
\newcommand{\een}{\end{eqnarray}}

\newcommand{\p}{\partial}
\renewcommand{\a}{\alpha}
\renewcommand{\b}{\beta}
\renewcommand{\t}{\tau}
\newcommand{\s}{\sigma}

\newcommand{\refb}[1]{(\ref{#1})}
\newcommand{\w}{\omega}
\newcommand{\bw}{\bar{\omega}}
\newcommand{\<}{\langle}
\renewcommand{\>}{\rangle}

\title{Tensionless Strings and Galilean Conformal Algebra}

\author{Arjun Bagchi$^{\, a,b}$ 
\\
$\;$ $\,$ $^a \,$Center for Theoretical Physics, Massachusetts Institute of Technology \\
$\;$ $\,$ 77 Massachusetts Avenue, Cambridge. MA 02139, USA. \\

 $\;$ $\,$$^b \,$School of Mathematics and the Maxwell Institute of Mathematical Sciences \\
$\;$ $\,$University of Edinburgh. Edinburgh EH9 3JZ, UK. \\

$\;$\email{arjun.bagchi@ed.ac.uk, abagchi@mit.edu}
}

\abstract{We find an intriguing link between the symmetries of the tensionless limit of closed string theory and the 2-dimensional Galilean Conformal Algebra (2d GCA). 2d GCA has been discussed in the context of the non-relativistic limit of AdS/CFT and more recently in flat-space holography as the proposed symmetry algebra of the field theory dual to 3d Minkowski spacetimes. It is best understood as a contraction of two copies of the Virasoro algebra. In this note, we link this to the tensionless limit of bosonic closed string theory. We show how it emerges naturally as a contraction of the residual gauge symmetries of the tensile string in the conformal gauge. We also discuss a possible ``dual" interpretation in terms of a point-particle like limit.}

\preprint{MIT-CTP-4445, EMPG-13-02}

\begin{document}

\baselineskip 3.5ex

\section{Introduction}
\noindent
{
Conformal field theory (CFT) has been one of the main tools of theoretical physics in recent years. After starting life as a mathematical curiosity as an invariance of classical equations like the Maxwell equations, conformal symmetry has now become central to understanding the nature of quantum field theories through renormalisation group fixed points \cite{yellow}. 
Our modern understanding of critical phenomena is based on CFTs. Techniques of CFTs are vitally important to string theory where conformal symmetry manifests itself as a gauge symmetry on the worldsheet, viz. the Weyl invariance of the classical string action. Over the last decade, CFTs have also been used extensively to understand holography and the gauge/gravity correspondence \cite{Maldacena:1997re}. 

The conformal symmetry on the closed string world sheet is 2 dimensional and it is in these dimensions that CFTs are at their most powerful \cite{Belavin:1984vu}. The underlying symmetry algebra is enhanced to two copies of the infinite dimensional Virasoro algebra. The infinite dimensional nature of the Virasoro algebra enables us to fix an enormous amount of the theory under question just by symmetry arguments. For example, one does not need a Lagrangian formulation to understand important quantities like the correlation functions which are highly constrained and at times even fixed by symmetries. 

Recently, another symmetry algebra called the Galilean Conformal Algebra (GCA) has been unearthed in seemingly very different contexts. It was first discussed in relation to the non-relativistic limit of the AdS/CFT correspondence \cite{Bagchi:2009my}. The finite dimensional GCA was constructed as a In{\"{o}}n{\"{u}}-Wigner contraction of the relativistic conformal algebra. It was found that the GCA could be given an infinite dimensional lift for all space-time dimensions. This also turned out to be related to the symmetries of non-relativistic hydrodynamic equations \cite{russian}. In two dimensions, the infinite GCA was shown to arise very naturally from the contraction of two copies of the infinite Virasoro symmetry \cite{Bagchi:2009pe}. Given the vast body of literature on the techniques of 2d CFTs, this provided a very useful route to understanding similar quantities in the context of the 2d GCA.  

The asymptotic symmetries of flat space at null infinity is governed by infinite dimensional Bondi-Metzner-Sachs algebras in 4d \cite{Bondi:1962px, Sachs:1962zza} and in 3d \cite{{Barnich:2006av}}. In a more recent and intriguing development, it was found that the 2d GCA was isomorphic to BMS$_3$ and generalisations of the GCA in 3d were the symmetries of flat space in 4d. This connection was dubbed the BMS/GCA correspondence \cite{Bagchi:2010eg}. Given that the 2d GCA is a contraction of the Virasoro algebra, this meant that the symmetry structure of flat space could be understood as a limit of the symmetry structure of AdS \cite{Bagchi:2012cy, Barnich:2012aw}. This provided support for the idea that flat holography could indeed be understood as a limit of usual AdS/CFT \footnote{The contraction that is relevant to flat holography is actually ultra-relativistic as opposed to the non-relativistic case we described earlier. The magic of two dimensions means that the contracted algebras in both these limits are isomorphic.}. An effort to build flat-space holography following this line of thought has met with recent interesting successes \cite{Bagchi:2012yk}--\cite{Barnich:2012rz} and is currently being actively pursued. We should add that the efforts so far have been principally confined to understanding the 3d bulk and 2d boundary theory. 

In this note, we uncover yet another realisation of the GCA, this time in a limit of string theory. Our focus in this note is on the 2d GCA. We find a realisation of 2d GCA as the residual gauge symmetry of the tensionless string in the analogue of the conformal gauge. We interpret why the 2d GCA arises naturally in the tensionless limit of the closed string and plays the same role that Virasoro $\otimes$ Virasoro plays for the tensile closed string theory. 

The tensionless limit of string theory has been of interest ever since it was discussed by Schild \cite{Schild} nearly four decades ago. The limit is expected to probe the properties of string theories at very high energies much like the massless limit of particle theory probes the short-distance behaviour of ordinary field theories \footnote{There have been numerous papers which have addressed this subject over the years. We would like to mention that we are not attempting a historical account here and would only refer to papers which are of direct relevance to our work. We apologise in advance for any accidental oversight of potentially related older work.}. The spectrum of tensile string theory includes an infinite tower of massive particles of arbitrary spin. In the tensionless limit, all these particles become massless and the theory is thought to exhibit higher-spin symmetry. There has been recent renewed interest in this subject with the development of new higher spin holographic dualities linking Vasiliev higher spin theories in the bulk \cite{Vasiliev:2004qz} to theories with enhanced symmetry on the boundary. These are dualities which link Vasilev theory in AdS$_4$ to the O(N) vector model in 3d \cite{Klebanov:2002ja, Sezgin:2002rt} and Vasiliev theory on AdS$_3$ to the large N limit of $W_N$ minimal models in 2d \cite{{Gaberdiel:2010pz}}. It has also long been believed that the tensionless limit of type IIB string theory in AdS$_5 \times$S$^5$ should reduce to a higher spin gauge theory \cite{Witten-talk, Sundborg:2000wp}. 

While a large body of recent literature has emerged building on the higher-spin holographic dualities (see e.g. \cite{Gaberdiel:2012uj} for the developments in the 3d-2d duality), there has been limited progress in understanding concretely the link between tensionless string theory and these higher spin structures\footnote{See, however, \cite{Sagnotti:2003qa}. Near the end of this paper, we discuss the relation of our work to the literature following \cite{Sagnotti:2003qa}.}. This, we believe, is due to a lack of structure in the study of tensionless strings. In the usual tensile string theory, the methods of 2d conformal field theory were instrumental in helping the community achieve its current understanding. This was pioneered by development of the organisation principle in the seminal work by Friedan, Martinec and Shenker \cite{Friedan:1985ge}. We believe that the methods of 2d GCAs can have a similarly useful role to play in understanding the various aspects of tensionless strings.  

The fact that the 2d GCA emerges as a contraction of the Virasoro algebra has been exploited in \cite{Bagchi:2009pe} to show that one can learn many things about the field theories governed by the GCA (we shall henceforth call them Galilean Conformal Field Theories or GCFTs) by looking at proper limits of the corresponding answers from 2d CFTs. With the details of the limiting procedure at our disposal, we can look to construct all the tools needed to build the theory of tensionless strings in the same spirit as the 2d CFT was used to build the theory of tensile strings in the conformal gauge. 

In this note, we outline the first steps to this programme. In Sec \ref{sec2}, after a review of the construction of the action and symmetries of the tensionless limit of the closed bosonic string following \cite{Isberg:1993av}, we show how the symmetries arise as a contraction of the symmetries of the tensile string. We also construct the energy-momentum tensor of the GCFT as a first step to understanding the spectrum of the tensionless string in this gauge.  In Sec \ref{sec3}, we point out the similarities to the point-particle contraction and propose a ``duality" between the two extreme pictures. We further substantiate our claim by looking at the theory on a cylinder. In Sec \ref{sec4}, we comment on possible links to flat-space holography and conclude with some discussions about future directions.}

\section{Tensionless Strings}\label{sec2}

\subsection{Classical Tensionless Closed Strings}

We start off with a description of the classical theory of tensionless strings following \cite{Isberg:1993av}. We would focus on the symmetries of the Nambu-Goto action in the tensionless limit. 
\be{}
S = -T \int d^2 \xi \sqrt{-\det \gamma_{\alpha \beta}} 
\ee
where $\xi_a$ are world volume coordinates, $T$ is the tension and $\gamma_{\alpha \beta}$ is the induced metric 
\be{} 
\gamma_{\alpha \beta} = \p_\alpha X^m \p_\beta X^n \eta_{mn}.
\ee
$X^n$ are the spacetime co-ordinates of the string and $\eta_{mn}$ is the flat background metric. 
The generalised momenta derivable from the above action is given by 
\be{}
P_m = T \sqrt{-\gamma} \gamma^{0 \alpha} \p_\alpha X_m
\ee
They satisfy the constraints
\be{}
P^2 + T^2 \gamma \gamma^{00}=0, \,\ P_m\p_{\alpha} X^m =0
\ee
which make up the Hamiltonian of the system
\be{}
\mathcal{H} = \lambda(P^2 + T^2 \gamma \gamma^{00}) + \rho^\alpha P_m\p_{\alpha} X^m .
\ee
Here $\lambda, \rho^\alpha$ are the variables conjugate to the constraints in the Dirac formulation. The action after integrating out the momenta becomes \cite{Isberg:1993av}
\be{act}
S = \frac{1}{2} \int d^2 \xi \, \frac{1}{2 \lambda} \bigg[ {\dot{X}}^2 -2 \rho^\alpha{\dot{X}}^m \p_\alpha X_m + \rho^a \rho^b \p_b X^m \p_a X_m - 4 \lambda^2 T^2 \gamma \gamma^{00}\bigg]
\ee
If we identify 
\be{}
g^{\alpha \beta} = \begin{pmatrix} -1 & \rho \\ \rho & -\rho^2 + 4 \lambda^2 T^2 \end{pmatrix}
\ee
we can recast the action \refb{act} in the familiar Weyl invariant form
\be{}
S = -\frac{T}{2}  \int d^2 \xi \sqrt{-g} g^{\alpha \beta} \p_\alpha X^m \p_\beta X^n \eta_{mn}.
\ee
The tensionless limit can be taken at various steps in the above. The point to keep in mind is that the metric density $T \sqrt{-g} g^{\alpha \beta}$ would degenerate  in the limit. One can replace this by a rank one matrix that can be written as $V^\alpha V^\beta$ where $V^\alpha$ is a vector density
\be{}
V^\alpha \equiv \frac{1}{\sqrt{2} \lambda} (1, \rho^a)
\ee
The action in the $T \to 0$ limit then becomes 
\be{acti}
S = \int d^2 \xi \,\ V^\alpha V^\beta \p_\alpha X^m \p_\beta X^n \eta_{mn}.
\ee
The action of the tensionless string is invariant under world-sheet diffeomorphisms. And as usual, one needs to fix a gauge.  
It is particularly useful to look at the tensionless action in the analogue of the conformal gauge for the tensile string. 
\be{conf-gauge}
V^\alpha = (v, 0)
\ee
where $v$ is a constant. Just as in the tensile case, there is a residual symmetry that is left over after this gauge fixing. This is going to be the primary focus of our note. In the tensile case, the residual symmetry in the conformal gauge is two copies of the Virasoro algebra for the closed string. This infinite dimensional symmetry structure has been central to understand the theory of usual tensile strings. Following \cite{Isberg:1993av}, we first construct the form of the residual symmetries of the tensionless string in this gauge. The residual symmetries take the form
\be{}
\delta \xi^\a = \lambda^\a, \quad \lambda^\a = (f^{\prime}(\sigma) \tau + g (\sigma), f(\sigma))
\ee
where $f,g$ are arbitrary functions of $\sigma$. Defining the generators 
\be{}
L(f) = f^{\prime}(\sigma) \tau \p_\tau + f(\sigma) \p_\sigma, \quad M(f) = f(\sigma) \p_\tau
\ee
the symmetry algebra takes the form 
\ben{}
&&[L(f_1),L(f_2)] = L(f_1f^{\prime}_2 - f_2f^{\prime}_1), \qquad [M (g_1), M (g_2)]= 0,\crcr
&&[L(f),M(g)]=M(fg^{\prime} - gf^{\prime}).
\een
Fourier expanding $f, g$ as $f(\sigma) = \sum a_n e^{i n \sigma}$,  $g(\sigma) = \sum b_n e^{i n \sigma}$ we get
\ben{LM} 
L(f) &=& \sum_n a_n e^{in \sigma} (\p_\sigma + i n \tau \p_\tau) = - i \sum_n a_n L_n, \\
M(g) &=& \sum_n b_n e^{in \sigma} \p_\tau = - i \sum_n b_n M_n
\een
The symmetry algebra which is the residual symmetry in the analogue of the conformal gauge of the tensionless string then in terms of the Fourier modes defined above takes the form:

\ben{gca2d}
[ L_m, L_n] &=& (m-n) L_{m+n} + C_1 m(m^2 -1) \delta_{m+n,0},  \quad [M_m, M_n] =0. \cr
[L_m, M_n] &=& (m-n) M_{m+n} + C_2 m(m^2 -1) \delta_{m+n,0}.
\een
Here $C_1, C_2$ are the most general central charges that are consistent with Jacobi identities. \cite{Isberg:1993av} mentions that in the simplest limit both of these would be zero. We recognise \refb{gca2d} as the Galilean Conformal Algebra in two dimensions \cite{Bagchi:2009pe}. 2d GCA is known to arise from a contraction of two copies of the Virasoro algebra. In the next sub-section, we discuss how this contraction is very naturally connected to the tensionless limit of string theory.

\subsection{Tensionless limit and the emergence of 2d GCA.}
For the tensile string, two copies of the Virasoro algebra arise as the residual symmetry in the conformal gauge $g_{\a \b} = e^{\phi} \eta_{\a \b}$. We shall denote the Virasoro algebra of the tensile string as $\L_n, \bL_n$:
\ben{}
[ \L_m, \L_n] &=& (m-n) \L_{m+n} + \frac{c}{12} m(m^2 -1) \delta_{m+n,0}  \crcr
[\L_m, \bL_n] &=& 0, \quad  [ \bL_m, \bL_n] = (m-n) \bL_{m+n} + \frac{\bar{c}}{12} m(m^2 -1) \delta_{m+n,0}
\een
The world-sheet of the tensile closed string is a cylinder. The residual gauge symmetry is thus best expressed as 2d conformal generators on the cylinder. We consider the following vector fields
\be{lbl}
\L_n = i e^{in\w}\p_{\w}, \quad  \bL_n =  i e^{i n\bw}\p_{\bw}
\ee
where $\w = \t + \s$, $\bw = \t - \s$. These vector fields generate the centre-less Virasoro algebra. 

Now, the tensionless limit of string theory is the limit where the length of the string becomes infinite. For the co-ordinates on the world-sheet this is a limit of $\s \to \infty$. Since the ends of a closed string are identified, this limit is best viewed as a limit where $(\s \to \s, \t \to \e \t, \e \to 0)$. We shall call this the Ultra-Relativistic (UR) limit. The individual vector fields \refb{lbl} are singular in the limit. We shall thus be working with linear combinations of them which behave well under the contraction. Let us define
\be{ultra-lim}
L_n =\L_n - \bL_{-n}, \quad M_n =\e (\L_n + \bL_{-n}).
\ee
So, in the cylinder co-ordinates, before taking the limit the generators take the following form:
\be{} 
\L_n - \bL_{-n} = i e^{in\s} \left(i\sin{n\t}\, \p_\t +  \cos{n\t}\, \p_\s \right), 
\quad \L_n + \bL_{-n} = i e^{in\s} \left( \cos{n\t}\, \p_\t + i \sin{n\t} \,\p_\s \right) 
\ee 
Now we perform the contraction by taking the scaling 
\be{newscal} 
\t \to \e \t, \s \to \s, \,\ \mbox{where} \,\ \e\to 0.  
\ee 
The new vector fields $(L_n, M_n)$ are
well-defined in the $\e \to 0$ limit and are given by 
\be{gen} 
L_n = ie^{in\s}( \p_\s+ in\t \p_\t), \quad M_n =  i  e^{in\s} \p_\t. 
\ee
We see that these are identical to the form of the generators \refb{LM} that we had derived earlier following \cite{Isberg:1993av}. We should also point out that these are also the form of the generators that are encountered in the flat-limit of the AdS asymptotic symmetry group, which is an ultra-relativistic contraction of the two copies of the Virasoro algebra \cite{Bagchi:2012cy}. 

With the relativistic central charges $c, \bar{c}$ included in the Virasoro algebra, this leads to the 2d GCA
\ben{2dgca}
[ L_m, L_n] &=& (m-n) L_{m+n} + C_1 m(m^2 -1) \delta_{m+n,0},  \quad [M_m, M_n] =0. \cr
[L_m, M_n] &=& (m-n) M_{m+n} + C_2 m(m^2 -1) \delta_{m+n,0},
\een
Here the central charges have the expression $C_1 = \frac{c - \bar{c}}{12}$, $C_2 = \frac{\e ( c + \bar{c} )}{12}$. For the case where the original CFT has $c = \bar{c}$, $C_1=0$. Also for finite central charges $c, \bar{c}$, we would end up with $C_2=0$ as well.

\subsection{Central Charges and Critical Dimensions}

The central term in the Virasoro algebra of the closed bosonic string theory points to the existence of the conformal anomaly which leads to the unique dimension in which the string needs to propagate in order to be consistent. We have noticed above that for the residual symmetries of the tensionless string, we have $C_1=C_2=0$. The absence of central terms in the algebra would point to the fact that tensionless strings would be consistent in any number of dimensions and this absence of a critical dimension was earlier noted in \cite{Lizzi:1986nv} and subsequently by others. 

Let us elaborate further. If we want to look at the ultra-high energy behaviour of a particular consistent string theory, the central terms in the two copies of the parent Virasoro algebra would have to be equal (for the vanishing of the diffeomorphism anomaly) and finite. In that case, the 2d GCA, which would govern the tensionless limit of this particular string theory, would have vanishing central terms. This particular tensionless theory, on its own, would be consistent in any number of dimensions. This is possibly a satisfying outcome. One would not want a particular limit of a theory to give an additional constraint on the dimension in which the theory would be consistent other than perhaps the original critical dimension. 

On the other hand, we might be interested in a theory of tensionless strings without bothering about where the theory came from. In general, this could have non-zero $C_1$ and $C_2$. If we insist on understanding this again as arising from the limit of some theory, this parent theory would have pathologies. For generating a non-zero $C_1$, one could need to start with a parent with a diffeomorphism anomaly. To get a non-zero $C_2$, one needs to start with a theory having central charges that would scale like $1/\e$, which means that the parent needs to have an infinite number of world-sheet scalar fields. Obviously neither would be a consistent tensile string theory. 

It would be of interest to construct the analogue of the Weyl anomaly calculation for the 2d GCA to compare and contrast with the case of the Virasoro algebra. A first step to that would be the construction of an energy momentum tensor for the 2d GCFT which is what we do in the following sub-section.

\subsection{GCA E-M tensor and Physical State condition}

One of the central objects in the study of conformal field theories is the Energy-Momentum (E-M) tensor. The generators of the Virasoro algebra form the modes of the E-M tensor. In the construction of the spectrum of the tensile string theory in the conformal gauge in flat space, after quantising the world sheet theory as a theory free scalar fields, one imposes the constraint of the vanishing of the equation of motion of the metric which has been fixed to be flat. In an operator language this reduces to the statement that the physical states vanish under the action of the modes of E-M tensor. This forms the basis of decoupling unphysical negative norm states from the Hilbert space. 

In order to build to this goal for the tensionless string, we want to understand how to construct the mode expansion of the E-M tensor for the 2d GCFT as a limit of the construction of 2d CFT \footnote{The construction here is similar to the one outlined previously in \cite{Bagchi:2010vw}, but the details are different as the limit is an ultra-relativistic one here in contrast to the non-relativistic one discussed in \cite{Bagchi:2010vw}.}. For this, we start with the energy-momentum tensors for the Virasoro algebra on the cylinder. 
\ben{}
T_{cyl} &=& z^2 T_{plane} - \frac{c}{24} = \sum_{n} \L_n z^{-n} - \frac{c}{24} =  \sum_{n} \L_n e^{-in\w} - \frac{c}{24}; \nonumber \\ \bar{T}_{cyl} &=& \sum_{n} \bL_n e^{-in\bw} - \frac{\bar{c}}{24}
\een
Now, the generators are 
\be{}
\L_n = L_n + \frac{1}{\e} M_n \quad \bL_n = - (L_{-n} - \frac{1}{\e} M_{-n} )
\ee
We also need to take a contraction of $(\t \to \e \t , \, \s \to \s, \, \e \to 0)$ with $\w, \bw = \t \pm \s$.
 
\ben{}
&&T_{cyl} = \sum_{n} \bigg[( L_n  - in\t M_n) e^{-in\s} + \frac{1}{\e} M_n e^{-in\s} \bigg] - \frac{c}{24} \\
&& \bar{T}_{cyl} = \sum_{n}[- ( L_n  - in\t M_n) e^{-in\s} + \frac{1}{\e} M_n e^{-in\s} \bigg]- \frac{\bar{c}}{24}
\een
So, the energy momentum tensors in the limit of $\e \to 0$ are defined by
\ben{}
&&T_{(1)} = \lim_{\e \to 0} \bigg( T_{cyl} - \bar{T}_{cyl} \bigg) = \sum_{n} ( L_n  - in\t M_n) e^{-in\s} - \frac{C_1}{2} \\
&&T_{(2)} = \lim_{\e \to 0} \e \bigg( T_{cyl} + \bar{T}_{cyl} \bigg) = \sum_{n} M_n e^{-in\s} - \frac{C_2}{2}
\een
For looking at the spectrum of tensionless strings, we realise that the physical spectrum would be restricted by a constraint of the form 
\be{}
\< \mbox{phys} | T_{(1)} | \mbox{phys}' \> = 0, \quad \< \mbox{phys} | T_{(2)} | \mbox{phys}' \> = 0
\ee
which boils down to 
\be{}
L_n | \mbox{phys} \> = 0, \quad M_n | \mbox{phys}\> = 0 \quad \mbox{for $n>0$}
\ee
There is a subtlety which one needs to keep in mind here. The energy momentum tensor of any system is best defined as the variation of the action under changes of the metric. As we shall see later, the metric on the worldsheet in this tensionless limit is degenerate. This is reflected in the above analysis as we see that the E-M tensor $T_{(2)}$ needs to be defined with a factor of $\e$ to render it finite. This would be associated with the component of the world-sheet metric which vanishes in the limit. 

\newpage

\section{The ``Dual" frame}\label{sec3}

\subsection{The other contraction}

For the purposes of this note, we have confined ourselves to a Euclidean world-sheet and we shall continue to do so. As far as local properties are concerned, the 2d CFT of the tensile string treats $\t$ and $\s$ on an equal footing. So it should also treat contractions of the two directions equally, i.e. a contraction $(\t, \s) \to (\e \t, \s)$ or the other viz. $(\t, \s) \to (\t, \e \s)$ should yield the same local physics. This implies that the symmetry structures, especially the symmetry algebra should be the same in these two different limits. We have already seen how the GCA emerges from the Virasoro algebra when one takes $(\t, \s) \to (\e \t, \s)$ on the generators of the Virasoro algebra on the cylinder. Now, let us consider the other contraction $(\t, \s) \to (\t, \e \s)$. We shall call this the Non-Relativistic (NR) limit. We define
\be{non-lim}
\tilde{L}_n = \L_n + \bL_n, \quad \tilde{M}_n = \e(\L_n - \bL_n)
\ee
where $\L_n, \bL_n$ are defined by \refb{lbl}. Taking the limit $\e \to 0$, it can be shown that the expressions for the limiting generators take the form
\be{tilde} 
\tilde{L}_n = ie^{in\t}( \p_\t+ in\s \p_\s), \quad \tilde{M}_n =  i  e^{in\t} \p_\s. 
\ee
These close to form the 2d GCA. It is also evident that the expressions for the generators \refb{tilde} and \refb{gen} go into each other by a $\s \leftrightarrow \t$ interchange. It should be noted however that the central charges $C_1$ and $C_2$ interchange roles and demanding that the parent CFT has equal left and right moving central charges makes $C_2=0$ and does not affect $C_1$. We would like to interpret this as a local versus global feature and point to the fact that the global features of the theory would dictate what the central charge is. Since we have exchanged a compact direction ($\s$) with a non-compact one ($\t$), one would not expect the global features to match. 

From the above description, it is clear that the tensionless limit of the closed string, where the string becomes extremely long, should have features very similar to the other extreme limit, the point-particle limit of string theory. This is a striking feature and is explicitly exemplified by the equations of motion of the tensionless string in this analogue of the conformal gauge. 
From the action \refb{acti}, one can derive the field equations. These are \cite{Isberg:1993av}
\be{eom}
V^\b \gamma_{\a \b} = 0, \quad \p_\a(V^\a V^\b \p_\b X^m) =0 
\ee
Let us for now focus on the second equation in \refb{eom}. This is best interpreted in the ``conformal" gauge \refb{conf-gauge} where it becomes 
\be{}
{\p_\t}^2 X^m = 0, \quad ({\p_\t {X}})^2 = \p_\t X \p_\s X = 0.
\ee
This implies that the string in this limit behaves like a collection of massless point particles. This ties in nicely with the ``dual" picture of the equivalence of the two contractions. The point-particle like nature of the tensionless is further illustrated by the emergence of a space-time conformal symmetry from the action. For details on this, the reader is referred to \cite{Isberg:1993av}. 

Let us return to the first of the equations of motion \refb{eom}. This implies that the matrix $\gamma_{\a \b}$ has an eigenvector with eigenvalue zero and hence is a degenerate matrix. The surface of propagation of the tensionless string is thus a null hypersurface. This is why the tensionless string was originally called the null string.  The fact that the metric defining the worldsheet of the tensionless string is degenerate is also something that one can understand from the symmetry structure, viz. it is a property of the 2d GCA. 

The Galilean Conformal Algebra is naturally associated with degenerate metrics. It has been well known that the process of contraction leading to the GCA is not a smooth limit on the metric of the spacetime. This is analogous to the behaviour of the Minkowski metric in the $c \to \infty$ limit ($c$ is the speed of light):
\be{}
\eta^{\mu\nu} = \mbox{diag} ( -\frac{1}{c^2}, 1,1, \ldots), \quad \eta_{\mu\nu} = \mbox{diag} ( -{c^2}, 1,1, \ldots)
\ee
This degeneration is well known and the way around this apparent singular spacetime description is to understand that the metric is not the correct dynamical variable to use in this context. One can either work in terms of the dynamical connection \cite{Misner:1974qy, Bagchi:2009my}, or in terms of the co-metric \cite{Brattan:2011my}. In case of the non-relativistic limit of flat-space, the manifold, called the Galilean manifold, is endowed with a fibre-bundle structure with the time axis $R_t$ as the base  and d-dimensional spatial fibres $R^d$ which are endowed by a spatial metric $\delta_{ij}$. 

As is obvious from the above discussion, the ultra-relativistic limit $c\to 0$ would work in a similar way with an exchange of roles of the spatial and temporal directions. This is the limit to look at for understanding flat space holography. We have mentioned that 3d flat-holography is linked with the 2d-GCA.  This feature of the degeneration of the metric was also something one encountered there. The dual field theory of flat space, governed by the symmetries of the 2d-GCA, would exist on the null boundary of flat space $\mathcal{I}^\pm$. The metric on this null surface is also obviously degenerate. 

In the light of the above discussion, it may be of interest to re-formulate the theory of tensionless strings in terms of the connections on the world-sheet instead of the metric and understand the emergence of the E-M tensors defined previously in terms of this structure.

\subsection{Equivalence of Contractions: Theory on a Torus}

We mentioned that on a Euclidean world-sheet the theory does not distinguish between time and space and hence should not distinguish between the two different contractions mentioned. The local properties should be the same and this is reflected by the symmetry algebra. But since one is trading a contraction in a compact direction with a contraction in a non-compact direction, there would be global properties which would differ between the two limits. This distinction is also lost when we take the time direction to be compact or we consider the theory on a torus. In this section, we present a non-trivial check in support of this statement. We demonstrate that for the theory on a torus, the two limits give the same asymptotic counting of states, i.e. reproduce the same Cardy-like formula for the 2d GCFT\footnote{The formulae have been derived earlier for the non-relativistic limit in \cite{hotta} and for the flat-limit in \cite{Bagchi:2012xr}. This section is an explicit demonstration of the equivalence of the two formulae.}.  

Our analysis would be a calculation inherently in the GCFT and the only input from the fact that the GCA is obtained as a limit of the two copies of the Virasoro algebra would be in identifying the modular invariance of the GCFT. We would assume that the partition function of the GCFT inherits a modular invariance from the parent 2d CFT it comes from. 

\paragraph{\underline{UR Limit}:} Here we first consider the limit we first discussed in Sec 2 where the GCA generators are mapped to the Virasoro generators by \refb{ultra-lim}. The states in the 2d GCFT are labelled by the eigenvalues of $L_0, M_0$ \cite{Bagchi:2009ca}.
\be{l0m0}
L_0 | h_L, h_M \> = h_L  | h_L, h_M \>, \quad M_0 | h_L, h_M \> = h_M  | h_L, h_M \>
\ee
In this limit, $(h_L, h_M)$ are mapped to the original eigenvalues of $\L_0, \bL_0$, $(h, \bar{h})$ by
\be{}
h_L = h - \bar{h}, \quad h_M = {\e} (h + \bar{h}).
\ee
In the analysis of the Cardy-like formula, we start with the CFT partition function and re-write it in the ``GCA-basis". 
\be{pf-cft}
Z_{\textrm{\tiny CFT}} = {\mbox{Tr}} \, e^{2 \pi \zeta \L_0} e^{-2\pi \bar{\zeta} \bL_0} =  \sum d_{{\textrm{\tiny CFT}}} (h, {\bar{h}}) e^{2 \pi i (\zeta h-{\bar{\zeta}} {\bar{h}})}= \sum d(h_{\textrm{\tiny L}}, h_{\textrm{\tiny M}}) e^{2 \pi i (\eta h_{\textrm{\tiny L}}+\frac{\rho}{\e} h_{\textrm{\tiny M}})} 
\ee
$\zeta, \bar\zeta$ are the modular parameters of the original 2d CFT. Above we have relabelled 
\be{etarho}
2\eta = \zeta + \bar \zeta. \quad 2\rho = \zeta - \bar \zeta
\ee
The S-transformation in the original CFT is $\zeta \to - \frac{1}{\zeta}$ and $\bar\zeta \to - \frac{1}{\bar \zeta}$.
This transforms to
\be{S-trans-gca}
\eta \to \frac{ \eta}{\rho^2 - \eta^2} \qquad  \rho \to - \frac{\rho}{\rho^2 - \eta^2}
\ee
We demand that the partition function of the parent CFT reduce to the GCFT partition function smoothly. 
This implies that $\rho$ has to scale for \refb{pf-cft} to stay finite in the limit. 
\be{}
Z_{\textrm{\tiny CFT}} \xrightarrow{\e\to0} Z_{\textrm{\tiny GCFT}} \Rightarrow \rho \to \e \rho
\ee
We note here that $\rho$ is the variable associated with $M_0$. $M_0 = \p_\t$ is the generator of world-sheet time translations and hence the world-sheet Hamiltonian. This is scaled in the limit and it necessitates the scaling of $\rho$ which is like a world-sheet temperature. The S-transformation of the GCFT which is inherited from the parent CFT \refb{S-trans-gca} thus reads:
\be{S-gca1}
(\eta, \rho) \to \left(- \frac{1}{\eta}, \frac{\rho}{\eta^2}\right)
\ee
The basic result would hinge on the invariance of quantity
\be{pf0}
Z^0_{\textrm{\tiny GCFT}} (\eta, \rho) = {\mbox{Tr}} \,\ e^{2 \pi i \eta (L_0 - \frac{C_1}{2})} e^{2 \pi i \rho (M_0 - \frac{C_2}{2})}= e^{\pi i (\eta {C_1} + \rho C_2)} Z_{\textrm{\tiny GCFT}}(\eta, \rho) 
\ee
under the inherited S-transformation of the GCFT. Here 
\be{pf}
Z_{\textrm{\tiny GCFT}} (\eta, \rho) = {\mbox{Tr}} \,\ e^{2 \pi i \eta L_0 } e^{2 \pi i \rho M_0} = \sum d(h_L, h_M) e^{2 \pi i \eta h_L} e^{2 \pi i \rho h_M}
\ee
In this limit, the modular invariance reads
\be{}
Z^0_{\textrm{\tiny GCFT}} (\eta, \rho) = Z^0_{\textrm{\tiny GCFT}} \bigg(- \frac{1}{\eta},\frac{\rho}{\eta^2}\bigg)
\ee
The main observation is now that we can translate this into statements for the partition function.
\be{}
Z_{\textrm{\tiny GCFT}} (\eta, \rho) = e^{2 \pi i \eta \frac{C_1}{2} } e^{2 \pi i \rho \frac{C_2}{2}} e^{- 2 \pi i (- \frac{1}{\eta} )\frac{C_1}{2} } e^{- 2 \pi i (\frac{\rho}{\eta^2}) \frac{C_2}{2}} Z_{\textrm{\tiny GCFT}} \bigg(- \frac{1}{\eta},\frac{\rho}{\eta^2}\bigg)
\ee
By doing an inverse Laplace transformation, one can find the density of states which was defined previously above in \refb{pf}.
\be{den}
d(h_L, h_M) = \int d \eta d \rho \,\ e^{2 \pi i {\tilde{f_1}}(\eta, \rho)} Z \bigg(- \frac{1}{\eta},\frac{\rho}{\eta^2}\bigg).
\ee
where
\be{}
{\tilde{f_1}}(\eta, \rho) =  \frac{C_1 \eta}{2} +  \frac{C_2 \rho}{2} + \frac{ C_1}{2\eta} - \frac{C_2 \rho}{2\eta^2} - h_L \eta - h_M \rho.
\ee
In the limit of large charges, the above integration \refb{den} can be performed by the method of steepest descents and the value of the integral is approximated by the value of the integrand when the exponential piece is an extremum. The saddle-point approximation used here is valid when one has an integrand with a rapidly varying phase and a slowly varying prefactor. So, one assumes that the partition function is slowly varying at the extremum. This can be checked. 
In the limit of large charges, the function ${\tilde{f_1}}(\eta, \rho)$ is approximated by:
\be{}
{f_1}(\eta, \rho) = \frac{C_1}{2\eta} - \frac{C_2 \rho}{2\eta^2} - h_L \eta - h_M \rho.
\ee
The extremum of this is evaluated and the value at the extremum is given by
\be{}
{f_1}^{max}(\eta, \rho) = - i \bigg( C_1 \sqrt{\frac{h_M}{2C_2}} + h_L \sqrt{\frac{C_2}{2h_M}} \bigg).
\ee
The Cardy-like formula for the GCA in this limit is given by
\be{cardy1}
S_{(1)} = \ln d(h_L, h_M) =  2\pi\bigg( C_1 \sqrt{\frac{h_M}{2C_2}} + h_L \sqrt{\frac{C_2}{2h_M}} \bigg).
\ee

\vspace{1cm}

\paragraph{\underline{NR Limit}:} We now look at the other limit \refb{non-lim} by which one can construct the GCA from the 2d CFT. In the contraction of this linear combination leads to states labelled in the same way as \refb{l0m0}, but now with a different identification of the eigenvalues in terms of the parent eigenvalues. 
\be{}
h_L = h + \bar{h}, \quad h_M= \e (h - \bar{h}).
\ee
Again we look at the partition function in the changed basis:
\be{pf-cft1}
Z_{\textrm{\tiny CFT}} = {\mbox{Tr}} \, e^{2 \pi \zeta\L_0} e^{-2\pi \bar{\zeta} \bL_0} =  \sum d_{{\textrm{\tiny CFT}}} (h, {\bar{h}}) e^{2 \pi i (\zeta h-{\bar{\zeta}} {\bar{h}})}= \sum d_1(h_{\textrm{\tiny L}}, h_{\textrm{\tiny M}}) e^{2 \pi i (\rho h_{\textrm{\tiny L}}+\frac{\eta}{\e} h_{\textrm{\tiny M}})} 
\ee
where $\eta, \rho$ are defined as before \refb{etarho}.
Demanding that this be finite in the limit $\e \to 0$ now means that $\eta$ needs to scale ($\eta \to \e \eta$) and the inherited modular transformation has the form
\be{S-gca2}
S: (\eta, \rho) \to \left( \frac{\eta}{\rho^2}, - \frac{1}{\rho} \right)
\ee
The analysis that leads to the Cardy-like formula in this case is identical to the one discussed earlier. We state the main results. The density of states takes the form 
\be{den1}
d_1(h_L, h_M) = \int d \eta d \rho \,\ e^{2 \pi i {\tilde{f_2}}(\eta, \rho)} Z_{\textrm{\tiny GCFT}} \bigg(\frac{\eta}{\rho^2}, - \frac{1}{\rho}\bigg).
\ee
where
\be{}
{\tilde{f_2}}(\eta, \rho) =  \frac{C_1 \rho}{2} +  \frac{C_2 \eta}{2} + \frac{ C_1}{2\rho} - \frac{C_2 \eta}{2\rho^2} - h_L \rho - h_M \eta.
\ee
Maximising the above function in the limit of large charges and then computing the integral \refb{den1} in the saddle point approximation and taking its logarithm we arrive at the Cardy-like formula in this limit. 
\be{}
S_{(2)}= \ln d_1(h_L, h_M) =  2\pi\bigg( C_1 \sqrt{\frac{h_M}{2C_2}} + h_L \sqrt{\frac{C_2}{2h_M}} \bigg).
\ee
We see that this is identical to \refb{cardy1}, the Cardy-like formula in the other limit. It is also obvious that the full analyses of the Cardy-like formulae in the two limits are identical with a $\eta \leftrightarrow \rho$ swap.

\subsection{Utility of Duality}
One of the great successes of string theory is the utilisation of its web of dualities which link the various string theories. Very complicated calculations can be rendered simple by mapping to a particular duality frame. The expectation is that the duality that we have proposed here between the two contractions of the world-sheet leading to the tensionless (UR) limit and the point-particle (NR) limit would have a similar usefulness. Let us mention one such instance here. 

We have seen that in the UR limit, the generators of the 2d GCA are a combination of creation and annihilation operators of the original CFT \refb{ultra-lim}. This is however not the case in the NR limit \refb{non-lim}. When one looks to understand the representation theory in terms of primary operators defined in the GCA in analogy with the constructions of the 2d CFT, if one is interested in a construction where the Virasoro primaries descend naturally to GCA primaries, it is thus much more useful to look at the NR limit as opposed to the UR limit.  

Before closing this section, let us mention another point. We have derived the Cardy formulae of the two limits by demanding there exist modular transformations inherited from the parent CFT. The two limits give two different theories which are governed by the symmetries of the 2d GCA. Let us call them GCFT$_1$ and GCFT$_2$. The contracted modular transformations e.g. S-transformation (\ref{S-gca1}) exists within GCFT$_1$ and similarly \refb{S-gca2} exists in GCFT$_2$. 
Usually, in string theory, when one considers say the one-loop amplitude, one integrates over the modulus of the torus, thus summing over world-sheet tori of all kinds linked by $SL(2,Z)$ transformations. We might be concerned that while looking at contracted geometries in the two limits, we should be summing over these solutions and not treated them as ``duals". This apprehension is however unfounded. The two contracted versions of the modular transformations in the two limits are linked by a $\eta \leftrightarrow \rho$ flip. This is not a transformation in the original $SL(2,Z)$ modular group. (It is actually a conformal transformation, but not a global one.) So when looking at 1-loop amplitudes, we should not be summing over these two different classes of contracted tori. 


\section{Comments and Future directions}\label{sec4}

\subsection{Connections to Flat-space Holography}
The spacetime symmetry, or in more precise words the asymptotic symmetry group of AdS$_3$ is two copies of the infinite dimensional Virasoro algebra. As is well known, and as we have mentioned earlier in this note, 2d-conformal symmetry also arises in the world sheet of the tensile string theory as a residual symmetry in the conformal gauge. It is natural to wonder if one could relate these two symmetry structures by looking at strings propagating is AdS$_3 \otimes$ X$^7$. The objective is to understand whether the symmetries of string theory can induce the symmetry structure of the spacetime in which it propagates. This was attempted in the seminal papers \cite{Giveon:1998ns, Kutasov:1999xu}. In the limited context of Type II B string theory on  AdS$_3 \otimes$ S$^3\otimes$ T$^4$, this is said to provide a proof of the AdS/CFT correspondence. 

The asymptotic symmetries of flat space in three dimensions at null infinity is the infinite dimensional BMS$_3$ algebra. This was recently shown to be isomorphic to the 2d GCA \cite{Bagchi:2010eg} and this has lead to an endeavour to build flat space holography as a limit of the usual AdS/CFT correspondence \cite{Bagchi:2012cy}--\cite{Barnich:2012xq}. We have seen in this note that the 2d GCA also arises as the residual symmetries of the tensionless string. In light of the above discussion of a ``derivation" of the AdS/CFT correspondence from world-sheet symmetries \cite{Giveon:1998ns, Kutasov:1999xu}, it is natural to ponder on similar consequences for flat holography. Given that the symmetries of the tensionless strings and 3d flat space match in the same way as that of the tensile strings and AdS$_3$ (and both are ultra-relativistic contractions), it is tempting to speculate that a construction similar to \cite{Giveon:1998ns, Kutasov:1999xu} for the tensionless strings may be a potential route to ``proving" our version of flat-holography. We hope to report on this issue in the near future.  

\subsection{Summary}

In this note, we have shown how the symmetries of the tensionless strings in the analogue of the conformal gauge arise as a contraction of the Virasoro symmetries of the tensile closed bosonic strings. We have stressed that the 2d GCA would play a vital role in the systematic construction of the various properties of the tensionless string, very similar to the one played by 2d CFTs in the case of the usual string theory. We have commented on the apparent lack of a critical dimension. We have taken a preliminary step in organising the tensionless spectrum by constructing the E-M tensor of the GCA in this limit. 

We have also proposed a duality between the tensionless limit and the point-particle limit by appealing to the symmetry of the contractions on an Euclidean world-sheet. This is reinforced by the old observation of the behaviour of the tensionless strings as a bunch of massless particles. The symmetry algebras in both limits are the same, viz. the 2d GCA.  We have furthered this claim of duality between the limits by looking at the theory on a torus and demonstrating that the analogue of the Cardy formula inherited by the 2d GCA in both limits is one and the same.

\subsection{Future Directions}
There are numerous avenues for immediate investigation following the observations that we have just made in this note. Let us list a few of them below.

As we have stressed, now with the organising principle in place, we should be able to derive the spectrum of the tensionless closed string in a covariant way analogous to the case of the tensile string theory. There have been previous attempts at this in the literature, mainly in a light-cone formulation. It would be of interest to match our results which would use the techniques of the 2d GCA to the results that already exist. It would also be good to understand how the spectrum of tensile closed strings goes over to the tensionless spectrum explicitly in the limit. 

The construction of the equivalent of the Weyl anomaly for the 2d GCA is an important problem. It would enable us understand whether there can be a critical dimension of the tensionless string when we analyse it without reference to any parent well-behaved string theory. 

There are obvious generalisations to which we should extend our analysis. The case of open-strings and D-branes is of prime importance. This may require some new insight as there we would have only one copy of the Virasoro algebra to work with and there is no obvious candidate of a non-trivial contracted algebra. It is possible that this would lead to a ``chiral"- truncation of the GCA down to its Virasoro sub-algebra. The extension to the closed superstring should be relatively straight-forward with the symmetry algebra being the contraction of the super-virasoro, which is called the 2d SGCA \cite{Bagchi:2009ke, Mandal:2010gx}.  

We should mention here that there exists an extensive body of work which needs re-examination in light of our findings in this note. This body of work along the lines of \cite{Sagnotti:2003qa} uses a contraction of the open-string algebra for a BRST quantisation of the tensionless open strings. We would like to remind the reader that it was the full 2d CFT which was responsible for the organisation principle of string theory and hence it is the closed string sector that needs to be understood first even in the context of the tensionless strings. What emerges as the symmetries of the closed tensionless string is not just two copies of the ``trivial" contraction of a Virasoro algebra but a limit on the linear combination of both copies of the Virasoro algebra in the 2d CFT which gives rise to the 2d GCA. We believe that the BRST quantisation of the tensionless string should be more correctly done by constructing the BRST charge of the 2d GCA. It would be interesting to see if our present findings are in keeping with the analysis of \cite{Sagnotti:2003qa}. If both approaches were to be correct, then it would seem that the bosonic open and closed strings behave very differently in the tensionless limit. 

We have also briefly commented on aspects of flat-holography that we are looking to address with a construction of an analogue of \cite{Giveon:1998ns, Kutasov:1999xu}. It would also be of interest to see whether looking at a tensionless limit of the original construction in AdS$_3$ \cite{Giveon:1998ns, Kutasov:1999xu} can shed some light on the emergence of the higher spin dualities in AdS. Here we would also like to remark that there has been interesting work in understanding string theory in the zero-radius limit of AdS (e.g. in \cite{Clark:2003wk}). This is equivalent to constructing a theory of tensionless strings in AdS of a certain fixed radius. The emergence of a space-time Virasoro algebra in this limit in AdS$_3$ from the world-sheet is heartening \cite{deMedeiros:2003hr} and provides hope that a more careful analysis could uncover the expected full higher spin symmetry. 

Of particular relevance to the construction of the higher spin symmetry in this context of tensionless strings is an analysis of WZW models at a critical level \cite{Lindstrom:2003mg}. Although there are significant steps taken towards this goal in \cite{Lindstrom:2003mg} and subsequently in \cite{Bakas:2004jq}, the analyses suffer from what the authors claim to be the singular nature of the CFT in the limit. We are of the opinion that the problems arise because the authors work in a domain where the original CFT$_2$ is not a valid symmetry structure and has to be replaced by the 2d GCA. The approach to the critical level in the WZW model would lead to a contraction of the Virasoro algebras. We are hopeful that a reconsideration of \cite{Lindstrom:2003mg, Bakas:2004jq} in the light of the symmetries of the 2d GCA would be the correct path to follow in this endeavour. We hope to report on these intriguing issues in the near future.

\newpage

\section*{Acknowledgements}
Several helpful discussions with Rajesh Gopakumar, Daniel Grumiller, Shailesh Lal, James Lucietti, Sunil Mukhi, S. Prem Kumar and especially Joan Simon are gratefully acknowledged. 

The author would like to thank the Center of Theoretical Physics, MIT for hosting him during the completion of this work. His stay at MIT is sponsored by an Early Career grant awarded jointly by the Scottish University Physics Alliance and the Edinburgh Partnership in Engineering and Mathematics. This work is also supported by Engineering and Physical Sciences Research Council (EPSRC). 



\begin{thebibliography}{999}

\bibitem{yellow}
    P.~Di Francesco, P.~Mathieu and D.~Senechal,
  ``Conformal Field Theory,''
{\it  New York, USA: Springer (1997)}.


\bibitem{Maldacena:1997re}
  J.~M.~Maldacena,
  ``The large N limit of superconformal field theories and supergravity,''
  Adv.\ Theor.\ Math.\ Phys.\  {\bf 2}, 231 (1998)
  [Int.\ J.\ Theor.\ Phys.\  {\bf 38}, 1113 (1999)]
  [arXiv:hep-th/9711200].
  
\bibitem{Belavin:1984vu} 
  A.~A.~Belavin, A.~M.~Polyakov and A.~B.~Zamolodchikov,
  ``Infinite Conformal Symmetry in Two-Dimensional Quantum Field Theory,''
  Nucl.\ Phys.\ B {\bf 241}, 333 (1984).
    

\bibitem{Bagchi:2009my}
  A.~Bagchi and R.~Gopakumar,
  ``Galilean Conformal Algebras and AdS/CFT,''
  JHEP {\bf 0907}, 037 (2009)
  [arXiv:0902.1385 [hep-th]].
  
\bibitem{russian}
 V.~N. Gusyatnikova and V.~A. Yumaguzhin,
 `` Symmetries and conservation laws  of navier-stokes equations,''
  Acta Applicandae Mathematicae {\bf 15} (January, 1989) 65--81.  
  

\bibitem{Bagchi:2009pe} 
  A.~Bagchi, R.~Gopakumar, I.~Mandal and A.~Miwa,
  ``GCA in 2d,''
  JHEP {\bf 1008}, 004 (2010)
  [arXiv:0912.1090 [hep-th]].

\bibitem{Bondi:1962px} 
  H.~Bondi, M.~G.~J.~van der Burg and A.~W.~K.~Metzner,
  ``Gravitational waves in general relativity. 7. Waves from axisymmetric isolated systems,''
  Proc.\ Roy.\ Soc.\ Lond.\ A {\bf 269}, 21 (1962).
  
  \bibitem{Sachs:1962zza} 
  R.~Sachs,
  ``Asymptotic symmetries in gravitational theory,''
  Phys.\ Rev.\  {\bf 128}, 2851 (1962).


\bibitem{Barnich:2006av} 
  G.~Barnich and G.~Compere,
  ``Classical central extension for asymptotic symmetries at null infinity in three spacetime dimensions,''
  Class.\ Quant.\ Grav.\  {\bf 24}, F15 (2007)
  [gr-qc/0610130].


\bibitem{Bagchi:2010eg} 
  A.~Bagchi,
  ``The BMS/GCA correspondence,''
  Phys.\ Rev.\ Lett.\  {\bf 105}, 171601 (2010)
  [arXiv:1006.3354 [hep-th]].

\bibitem{Bagchi:2012cy} 
  A.~Bagchi and R.~Fareghbal,
  ``BMS/GCA Redux: Towards Flatspace Holography from Non-Relativistic Symmetries,''
  JHEP {\bf 1210}, 092 (2012)
  [arXiv:1203.5795 [hep-th]].
  
  \bibitem{Barnich:2012aw} 
  G.~Barnich, A.~Gomberoff and H.~A.~Gonzalez,
  ``The Flat limit of three dimensional asymptotically anti-de Sitter spacetimes,''
  Phys.\ Rev.\ D {\bf 86}, 024020 (2012)
  [arXiv:1204.3288 [gr-qc]].

\bibitem{Bagchi:2012yk} 
  A.~Bagchi, S.~Detournay and D.~Grumiller,
  ``Flat-Space Chiral Gravity,''
  Phys.\ Rev.\ Lett.\  {\bf 109}, 151301 (2012)
  [arXiv:1208.1658 [hep-th]].

\bibitem{Bagchi:2012xr} 
  A.~Bagchi, S.~Detournay, R.~Fareghbal and J.~Simon,
  ``Holography of 3d Flat Cosmological Horizons,''
  arXiv:1208.4372 [hep-th].

\bibitem{Barnich:2012xq} 
  G.~Barnich,
  ``Entropy of three-dimensional asymptotically flat cosmological solutions,''
  JHEP {\bf 1210}, 095 (2012)
  [arXiv:1208.4371 [hep-th]].

\bibitem{Barnich:2012rz} 
  G.~Barnich, A.~Gomberoff and H.~A.~Gonzalez,
  ``BMS$_3$ invariant two dimensional field theories as flat limit of Liouville,''
  arXiv:1210.0731 [hep-th].


\bibitem{Schild} 
  A.~Schild,
  ``Classical Null Strings,''
  Phys.\ Rev.\ D {\bf 16}, 1722 (1977).
  
\bibitem{Karlhede:1986wb} 
  A.~Karlhede and U.~Lindstrom,
  ``The Classical Bosonic String In The Zero Tension Limit,''
  Class.\ Quant.\ Grav.\  {\bf 3}, L73 (1986).

\bibitem{Vasiliev:2004qz} 
  M.~A.~Vasiliev,
  ``Higher spin gauge theories in various dimensions,''
  Fortsch.\ Phys.\  {\bf 52}, 702 (2004)
  [hep-th/0401177].


\bibitem{Klebanov:2002ja} 
  I.~R.~Klebanov and A.~M.~Polyakov,
  ``AdS dual of the critical O(N) vector model,''
  Phys.\ Lett.\ B {\bf 550}, 213 (2002)
  [hep-th/0210114].
  
  \bibitem{Sezgin:2002rt} 
  E.~Sezgin and P.~Sundell,
  ``Massless higher spins and holography,''
  Nucl.\ Phys.\ B {\bf 644}, 303 (2002)
  [Erratum-ibid.\ B {\bf 660}, 403 (2003)]
  [hep-th/0205131].

\bibitem{Gaberdiel:2010pz} 
  M.~R.~Gaberdiel and R.~Gopakumar,
  ``An AdS$_3$ Dual for Minimal Model CFTs,''
  Phys.\ Rev.\ D {\bf 83}, 066007 (2011)
  [arXiv:1011.2986 [hep-th]].
  
\bibitem{Witten-talk}
  E.~Witten,
  in JHS/60: Conf. in Honor of John SchwarzÕs 60th Birthday, 
  California Institute of Technology, Pasadena, CA, USA, Nov. 3 Ð 4, 2001;
  http://theory.caltech.edu/jhs60/witten/1.html

\bibitem{Sundborg:2000wp} 
  B.~Sundborg,
  ``Stringy gravity, interacting tensionless strings and massless higher spins,''
  Nucl.\ Phys.\ Proc.\ Suppl.\  {\bf 102}, 113 (2001)
  [hep-th/0103247].

\bibitem{Gaberdiel:2012uj} 
  M.~R.~Gaberdiel and R.~Gopakumar,
  ``Minimal Model Holography,''
  arXiv:1207.6697 [hep-th].

\bibitem{Friedan:1985ge} 
  D.~Friedan, E.~J.~Martinec and S.~H.~Shenker,
  ``Conformal Invariance, Supersymmetry and String Theory,''
  Nucl.\ Phys.\ B {\bf 271}, 93 (1986).

\bibitem{Isberg:1993av} 
  J.~Isberg, U.~Lindstrom, B.~Sundborg and G.~Theodoridis,
  ``Classical and quantized tensionless strings,''
  Nucl.\ Phys.\ B {\bf 411}, 122 (1994)
  [hep-th/9307108].

\bibitem{Lizzi:1986nv} 
  F.~Lizzi, B.~Rai, G.~Sparano and A.~Srivastava,
  ``Quantization Of The Null String And Absence Of Critical Dimensions,''
  Phys.\ Lett.\ B {\bf 182}, 326 (1986).
  
  \bibitem{Bagchi:2010vw} 
  A.~Bagchi,
  ``Topologically Massive Gravity and Galilean Conformal Algebra: A Study of Correlation Functions,''
  JHEP {\bf 1102}, 091 (2011)
  [arXiv:1012.3316 [hep-th]].

\bibitem{Misner:1974qy} 
  C.~W.~Misner, K.~S.~Thorne and J.~A.~Wheeler,
  ``Gravitation,''
  San Francisco 1973, 1279p

\bibitem{Brattan:2011my} 
  D.~Brattan, J.~Camps, R.~Loganayagam and M.~Rangamani,
  ``CFT dual of the AdS Dirichlet problem : Fluid/Gravity on cut-off surfaces,''
  JHEP {\bf 1112}, 090 (2011)
  [arXiv:1106.2577 [hep-th]].
  
  \bibitem{Bagchi:2009ca}
  A.~Bagchi and I.~Mandal,
  ``On Representations and Correlation Functions of Galilean Conformal
  Algebras,''
  Phys.\ Lett.\  B {\bf 675}, 393 (2009)
  [arXiv:0903.4524 [hep-th]].


\bibitem{hotta} 
  K.~Hotta, T.~Kubota and T.~Nishinaka,
  ``Galilean Conformal Algebra in Two Dimensions and Cosmological Topologically Massive Gravity,''
  Nucl.\ Phys.\ B {\bf 838}, 358 (2010)
  [arXiv:1003.1203 [hep-th]].


  

\bibitem{Giveon:1998ns} 
  A.~Giveon, D.~Kutasov and N.~Seiberg,
  ``Comments on string theory on AdS(3),''
  Adv.\ Theor.\ Math.\ Phys.\  {\bf 2}, 733 (1998)
  [hep-th/9806194].

\bibitem{Kutasov:1999xu} 
  D.~Kutasov and N.~Seiberg,
  ``More comments on string theory on AdS(3),''
  JHEP {\bf 9904}, 008 (1999)
  [hep-th/9903219].


\bibitem{Sagnotti:2003qa} 
  A.~Sagnotti and M.~Tsulaia,
  ``On higher spins and the tensionless limit of string theory,''
  Nucl.\ Phys.\ B {\bf 682}, 83 (2004)
  [hep-th/0311257].

\bibitem{Bonelli:2003kh} 
  G.~Bonelli,
  Nucl.\ Phys.\ B {\bf 669}, 159 (2003)
  [hep-th/0305155].


\bibitem{Bagchi:2009ke} 
  A.~Bagchi and I.~Mandal,
  ``Supersymmetric Extension of Galilean Conformal Algebras,''
  Phys.\ Rev.\ D {\bf 80}, 086011 (2009)
  [arXiv:0905.0580 [hep-th]].

\bibitem{Mandal:2010gx} 
  I.~Mandal,
  ``Supersymmetric Extension of GCA in 2d,''
  JHEP {\bf 1011}, 018 (2010)
  [arXiv:1003.0209 [hep-th]].

\bibitem{Clark:2003wk} 
  A.~Clark, A.~Karch, P.~Kovtun and D.~Yamada,
  ``Construction of bosonic string theory on infinitely curved anti-de Sitter space,''
  Phys.\ Rev.\ D {\bf 68}, 066011 (2003)
  [hep-th/0304107].
  
\bibitem{deMedeiros:2003hr} 
  P.~de Medeiros and S.~P.~Kumar,
  ``Space-time Virasoro algebra from strings on zero radius AdS(3),''
  JHEP {\bf 0312}, 043 (2003)
  [hep-th/0310040].

\bibitem{Lindstrom:2003mg} 
  U.~Lindstrom and M.~Zabzine,
  ``Tensionless strings, WZW models at critical level and massless higher spin fields,''
  Phys.\ Lett.\ B {\bf 584}, 178 (2004)
  [hep-th/0305098].


\bibitem{Bakas:2004jq} 
  I.~Bakas and C.~Sourdis,
  ``On the tensionless limit of gauged WZW models,''
  JHEP {\bf 0406}, 049 (2004)
  [hep-th/0403165].



\end{thebibliography}
\end{document}